\documentclass{emulateapj}

\usepackage{amssymb}
\usepackage{amsmath}
\usepackage{epsfig}
\usepackage{graphicx}
\usepackage{natbib}

\slugcomment{accepted by ApJ Letters February 11, 2010}

\begin{document}

\title{
Minimum Radii of Super-Earths: Constraints from Giant Impacts
}
\author{Robert A. Marcus\altaffilmark{1,a}, Dimitar Sasselov\altaffilmark{1}, 
Lars Hernquist\altaffilmark{1}, Sarah T. Stewart\altaffilmark{2}
}
\affil{$^1$Astronomy Department, Harvard University, Cambridge, MA 02138, USA}
\affil{$^2$Department of Earth and Planetary Sciences, Harvard University, 
Cambridge, MA 02138, USA}
\email{$^a$rmarcus@cfa.harvard.edu}

\begin{abstract}
  The detailed interior structure models of super-Earth planets show
  that there is degeneracy in the possible bulk compositions of a
  super-Earth at a given mass and radius, determined via radial
  velocity and transit measurements, respectively. In addition, the
  upper and lower envelopes in the mass--radius relationship,
  corresponding to pure ice planets and pure iron planets,
  respectively, are not astrophysically well motivated with regard to
  the physical processes involved in planet formation. Here we apply
  the results of numerical simulations of giant impacts to constrain
  the lower bound in the mass--radius diagram that could arise from
  collisional mantle stripping of differentiated rocky/iron planets.
  We provide a very conservative estimate for the minimum radius
  boundary for the entire mass range of large terrestrial planets.
  This envelope is a readily testable prediction for the population of
  planets to be discovered by the Kepler mission.
\end{abstract}

\keywords{planets and satellites: formation --- planetary systems}

\section{Introduction}

Since the first confirmed detection of an extrasolar planet in 1995,
the catalog of known exoplanets has grown to surpass 400. Precision
Doppler shift discovery techniques have now identified more than a
dozen exoplanets in the mass range from about 2 to 10--15 Earth masses
($M_{\oplus}$). They are expected to be terrestrial in nature and
unlike the gas giant planets in our solar system. This new class of
planets has been collectively termed ``super-Earths''
\citep{Melnick:2001}. Recently two transiting super-Earths were
discovered, providing for the first time radii, in addition to masses.
One of them---CoRoT-7b \citep{Leger:2009, Queloz:2009}---has high
density and is likely rocky.  The other one---GJ1214b
\citep{Charbonneau:2009}---has low density and is likely water rich
and surrounded by a small hydrogen-rich envelope. Many more
super-Earths and measurements of their radii are expected from the
Kepler mission \citep{Kepler:2009}.

Theorists anticipate a rich diversity in the bulk composition and
internal structure of super-Earths, which is reflected in the broad
band of possible radii on a planetary mass--radius diagram.  The band
corresponds to the anticipated range of mean densities.  There are
four distinct types of materials that could make up a planet in this
regime: silicates, iron alloys, volatiles/ices, and hydrogen--helium
gas. The range of possible mixing ratios between these materials leads
to degeneracies in the determination of bulk composition from radius
and mass alone \citep{Sasselov:2008, Adams:2008}.  As shown by
\cite{Valencia:2007b}, in order to restrict the range of possible bulk
compositions, precise radii and masses (to 5\% and 10\%, respectively)
have to be complemented with knowledge of stellar abundance ratios
(e.g., Si/Fe) and physical constraints on the maximum fraction of
H$_2$O or iron in a planet. The latter constraints place limits on the
maximum and minimum possible radii for solid planets, respectively.

In this Letter, we consider the minimum possible radius a super-Earth
could have at a given mass. Since we are interested in the limiting
case, our discussion can be confined to rocky planets composed of iron
and silicates with no ices/water or hydrogen--helium gas layers. Under
physically plausible conditions around normal stars, planet formation
will lead to differentiated super-Earths with an iron core and a
silicate mantle, with the proportions of each determined by the local
Si/Fe ratio \citep{Grasset:2009}. The only way to significantly
increase the mean density requires removal of the silicate mantle
while preserving the iron core. The most efficient method to strip the
mantle is by giant impacts, which are common in the final stages of
planet formation \citep[e.g.,][]{Chambers:2004}.  Given the large
gravitational potential of super-Earth planets, we suspect that
complete mantle stripping is not possible, and unlike the case of
asteroids, pure iron super-Earths do not exist around normal stars. In
this Letter, we analyze the results from numerical simulations of
planet--planet collisions and determine a theoretical lower limit on
the planetary radii of super-Earths.  We anticipate that observations
by the Kepler mission will test our predictions.

The initial conditions for our investigation are dependent on an
understanding of planet formation. These results and observations
could, in turn, help constrain theories for planet formation.  In
recent years, \cite{Ida_LinI} and \cite{Mordasini:2009a} have applied
detailed models of planet formation to the generation of synthetic
populations. Such synthetic populations can then be compared to the
observed distribution of known exoplanets to place statistically
significant constraints on planet formation models. For example, the
analysis of \cite{Mordasini:2009a, Mordasini:2009b} shows that the
core accretion model of planet formation can reproduce observed
populations.  In this model, small planetesimals collide to form
larger planetary embryos ($\sim0.1M_{\oplus}$), the most massive of
which come to dominate accretion in a process known as runaway growth.
This stage is followed by an oligarchic stage in which protoplanets
become relatively isolated after consuming the surrounding
planetesimals.  At this stage, collisions between comparably sized
large bodies, giant impacts, become important and dominate the end
stages of the formation of terrestrial planets.  Giant impacts have
been extensively modeled in the planetary embryo size regime
\citep{Agnor_Asphaug:2004, Asphaug:2009}; however they had not been
studied extensively up to Earth size, with the exceptions of the Moon
forming impact (e.g., \citealt{Canup:2004}). Recently, the first study
focused on collisions between super-Earths determined the criteria for
catastrophic disruption and derived a scaling law for mantle stripping
\citep{Marcus:2009}.

\section{Method and Assumptions}

To address the question of physically plausible minimum radii for
super-Earths, we consider (1) the minimum radius based on cosmic
chemical abundances and (2) subsequent mantle stripping by giant
impacts.

The initial state of super-Earths is based on two assumptions,
following the discussion in \cite{Valencia:2007b}. The first is that
all super-Earth sized planets have undergone differentiation. The
second is that the relative elemental abundances are the same as those
in the solar system and the solar neighborhood. The first assumption
is not restrictive since all the terrestrial planets and large
satellites in the solar system are known to be differentiated.
Further, if any super-Earth planet were not fully differentiated at
the time of the impact considered here, it would be impossible to
preferentially remove lighter materials, and thus our constraint on
the minimum radius possible from collisional mantle stripping would
still hold. The second assumption should also be adequate for the
super-Earths that we expect to be discovered in the near future
(close-by and in the pre-selected targets of the Kepler mission).

\cite{Valencia:2007b} discuss the process by which cosmic elemental
abundances constrain an initial minimum radius for super-Earths, which
we summarize here.  Volatiles, silicates and metals condense at
different temperatures. As the stellar nebula cools, the most
refractory elements condense first.  First, silicates condense at
temperatures between 1750 and 1060~K, followed by the metals (e.g.,
Fe, Ni) between 1450 and 1050~K \citep{Petaev_Wood:2005}, depending on
the pressure in the nebula. Finally, H$_{2}$O and other ices condense.
We estimate the maximum iron core mass fraction by considering the
relative mass fraction of major elements in the solar nebula: H at
74\%, O at 1.07\%, Fe at 0.1\%, Si at 0.065\% and Mg at 0.058\%.  We
consider Si and Mg to be practically equally abundant. During the
condensation sequence (for pressures $<10^{-4}$ bars), Si will
condense before Fe. If Fe remains immiscible, the largest core is
attained at a mass ratio of Si/Fe$\sim0.6$. In this case, the mantle
is effectively MgO+SiO$_{2}$ so that the Si/Fe ratio can be used as a
proxy for the mantle-to-core mass fraction.

We contemplate processes in the late stages of planet formation that
influence the final state of a planet starting from embryos with
normal cosmic abundances (an iron mass fraction of about 0.33, which
is the value for Earth).  Any process that preferentially induces the
escape of light elements (e.g., solar wind, gravitational escape) will
deplete the planet from volatiles, including H$_{2}$O, and perhaps
silicates.  The major widespread process that could change the
mantle-to-core mass fraction dictated by elemental abundances is giant
impacts. The most effective collisions for increasing the mean density
of a planet are likely to be near-equal mass planet--planet
encounters, where both bodies are dry (composed of iron and
silicates). There would be few such planets, especially in the
super-Earth mass range, in any planet-forming disk. Hence, such
collisions would be rare occurrences, rather than multiple in the
history of a given planet.

Further assumptions in our calculations include the following: (1) the
mantle is stripped in a single, late giant impact, when the planet is
nearly fully accreted; (2) almost all of the mass remaining in the
post-impact planet is from the largest remnant (in other words, the
smaller fragments are not re-accreted); (3) super-Earths form only in
the mass range 1-15 $M_{\oplus}$ and beyond this upper limit, runaway
gas accretion causes the planet to become a gas giant
\citep{Ida_LinI}.  If super-Earths do form at masses larger than 15
$M_{\oplus}$, the physics of collisional mantle stripping would not
alter, and thus our results could be extrapolated beyond this limit.
The large relative velocities necessary to strip a significant portion
of a planet's mantle are most likely to occur early in the final stage
of planet formation, during which time there are many small planetary
embryos ($\sim0.1M_{\oplus}$). In $N$-body simulations of terrestrial
planet formation, this stage has been shown to produce impact
velocities as high as six times the mutual escape velocity
\citep{Agnor:1999}. However, such high impact velocities are still
rare, so most of the planetary embryos that are eventually
incorporated into a fully accreted super-Earth will not have suffered
mantle-stripping collisions, thus erasing the signature of a small
number of such impacts.

While relative impact velocities are expected to be around the mutual
escape velocity when only super-Earth mass planets remain, there need
only be a single impact event to leave a large remnant with a high
iron mass fraction.  Note that clearing the smaller fragments from the
orbit of this largest remnant, so that they would not be re-accreted,
would most likely require the presence of either the protoplanetary
disk or another large planet in the system. Because we are interested
in presenting a lower limit for the radii of super-Earths as a
function of mass, we believe these assumptions are justified and will
at worst result in an {\it underestimation} of the minimum of the
mass--radius relationship.

The calculations of mantle stripping presented in this Letter are
derived from simulations of head-on collisions between super-Earths.
Such low angle impacts are most effective at stripping mantle material
\citep{Benz:1988, Benz:2007, Marcus:2009}. The impact angle has little
effect on the efficiency of mantle stripping between impact angles of
0$^{\circ}$ (head-on) and about 30$^{\circ}$. In this regime, highly
disruptive impact events are possible. As the impact angle increases
beyond 30$^{\circ}$, there is a sharp transition, beyond which
collisions enter the ``hit-and-run'' regime \citep{Agnor_Asphaug:2004,
  Asphaug:2006, Asphaug:2009, Marcus:2009}, in which the projectile
and target emerge from the impact largely intact.
 
\section{Results}

Using the velocity-dependent catastrophic disruption criteria of
\cite{Stewart_Leinhardt:2009} and the scaling law for changes to the
iron-to-silicate ratio via disruption of bodies with an initial iron
mass fraction of 0.33 from \cite{Marcus:2009}, we derive the impact
velocity (in km~s$^{-1}$) necessary to produce a post-impact largest
remnant of a specified iron mass fraction $f_{Fe}$:
\begin{equation}
  V_{i} = 10.5(f_{Fe}-0.33)^{0.505} \left( \frac{(1+\gamma)^{2.4}}{\gamma} \right)^{\frac{1}{1.2}}M_{targ}^{\frac{1}{3}}. 
  \label{eq:vcrit}
\end{equation}
Here, $\gamma$ is the projectile-to-target mass ratio and $M_{targ}$
is in Earth masses. The mass of the corresponding largest remnant is
\begin{equation}
  M_{lr} = [-1.2(f_{Fe}-0.33)^{\frac{1}{1.65}} + 1]M_{targ}(1+\gamma).
  \label{eq:mlr}
\end{equation}
These equations are obtained by combining Equations (2)-(4) of
\cite{Marcus:2009}

\begin{figure}
\figurenum{1}
\begin{center}
\includegraphics[scale=0.6]{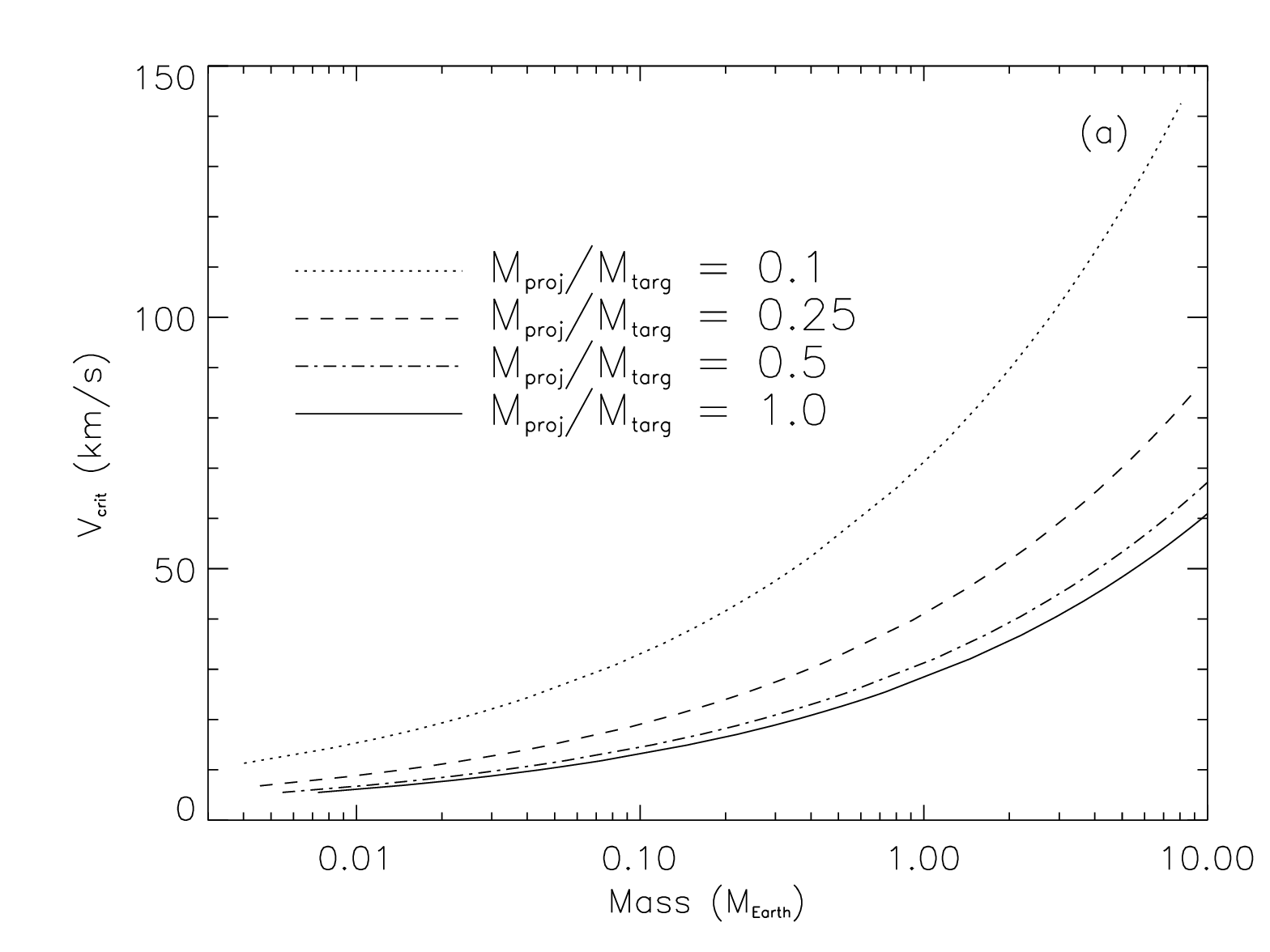}
\includegraphics[scale=0.6]{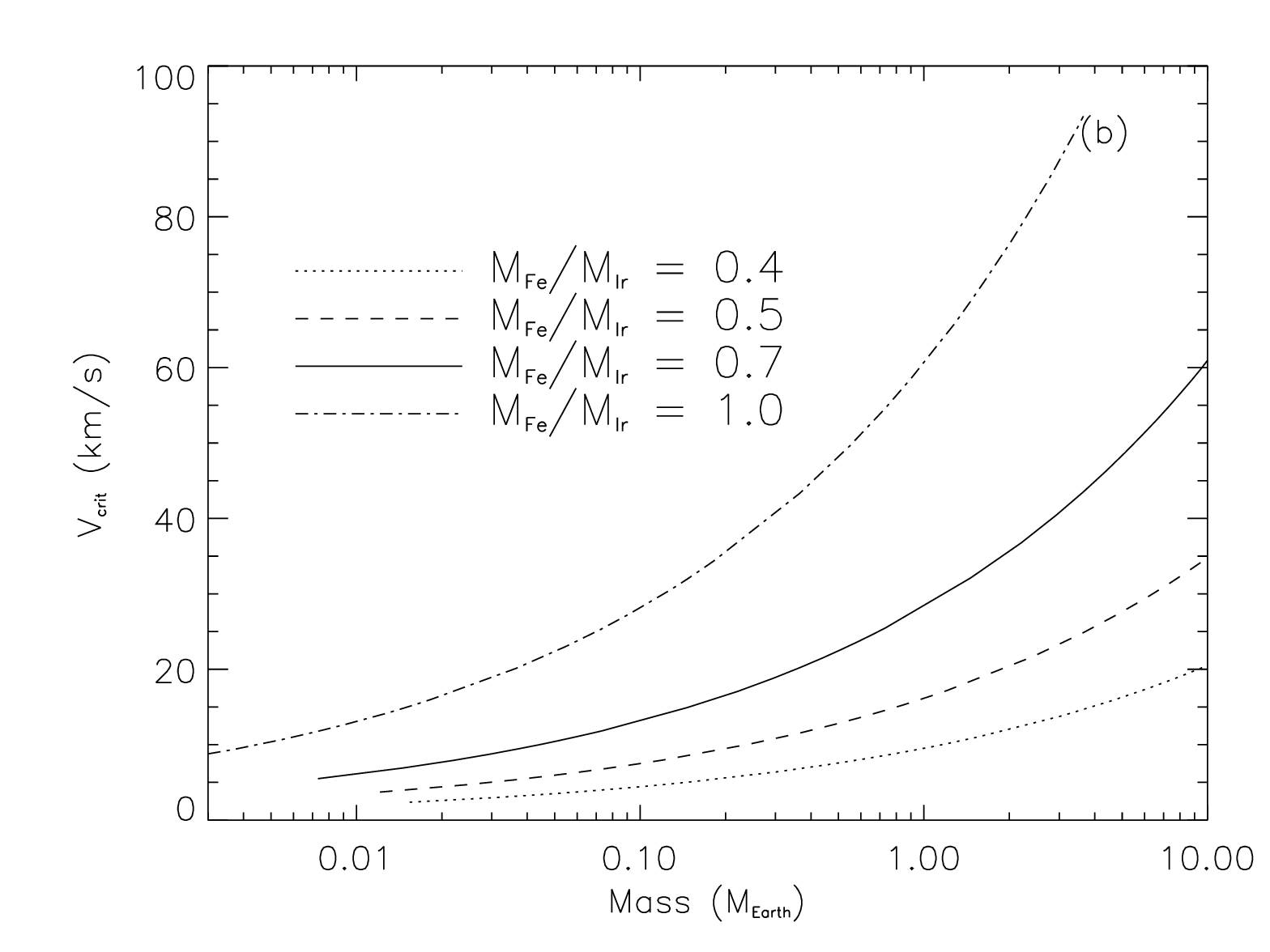}
\end{center}
\caption{Critical impact velocity required to obtain the specified
  post-impact iron mass fraction vs. mass of the largest impact
  remnant. (a) Largest remnant with iron mass fraction of 70\%
  (similar to Mercury) for various projectile-to-target mass ratios.
  (b) Largest remnant with various iron mass fractions for equal mass
  collisions at the lowest velocity for such an enrichment. Note that
  the solid line in both panels is for a 1:1 mass ratio collision
  yielding a largest remnant that is 70\% iron.}
\label{fig:vcrit}
\end{figure}

Figure \ref{fig:vcrit} presents this critical velocity as a function
of the mass of the largest remnant, which would be the observed
super-Earth. In Figure \ref{fig:vcrit}(a), the largest remnants have
$f_{Fe}=0.7$, making these planets super-Mercuries. For
projectile-to-target mass ratios $\lesssim$ 1/4, the critical velocity
rapidly exceeds 50~km~s$^{-1}$, making such collisions all but
impossible in the vicinity of 1 AU (the maximum possible impact
velocity at 1 AU around a solar-like star is $\sim$ 70~km~s$^{-1}$).
Such large impact velocities are comparable to the orbital velocity at
a location of 0.1 AU (for a 1$M_{\odot}$ star), the location at which
super-Earth planets may end Type I migration \citep{Masset:2006}.  For
masses close to 1$M_{\oplus}$ and projectile-to-target mass ratios
$\gtrsim$ 1/2, the critical velocities are $\lesssim$ 40 km~s$^{-1}$.
In Figure \ref{fig:vcrit}(b), the projectile-to-target mass ratio is
fixed at 1:1, the limiting (most destructive) case. The impact
velocities necessary to produce super-Earths with 40\%-50\% iron by
mass are 15-35 km~s$^{-1}$.

Figure \ref{fig:ironfrac} presents the iron mass fraction, $f_{Fe}$,
of the largest remnant after an impact as a function of the
projectile-to-mass ratio, impact velocity, and mass of the largest
remnant. The largest possible iron mass fraction resulting from mantle
stripping in a giant impact is clearly a function of the final mass of
the planet and the impact conditions. As expected, it is far easier to
remove mantle material from a lower mass super-Earth.

Next, the minimum mass--radius relationship for super-Earths can be
reconsidered based on the likelihood of the impact conditions
necessary to achieve a certain bulk density (based on the iron mass
fraction).  Figure \ref{fig:mass_radius} presents the radii of the
post-impact remnants as a function of the projectile-to-mass ratio and
the impact velocity.  The radii were calculated from the results of
the super-Earth internal structure models of \cite{Valencia:2006,
  Valencia:2007b}. Thus, these radii are not the radii of the largest
remnants immediately after the collision, at which time the largest
remnant consists of gravitationally reaccumulated debris (a rubble
pile).  Rather, these radii correspond to the radius of the
super-Earth with a core mass fraction given by $f_{Fe}$ long after
differentiation of the reaccumulated body and radiative loss of the
excess thermal energy from the collision. This corresponds to the time
at which we are most likely to observe the planet, given that
formation takes only $\sim$ 10 Myr of the planet's billion year-plus
lifetime. From Figure \ref{fig:mass_radius}, it is clear that even
given quite extreme impact conditions, with velocities of up to
80~km~s$^{-1}$, the minimum curve in the mass--radius diagram lies
well above the case for pure iron (lower black line), particularly for
more massive super-Earths.

Figure \ref{fig:mass_radius2} presents the mass--radius diagram for
super-Earths, with upper and lower envelopes (dotted lines) as
calculated by \cite{Fortney:2007}.  We add to this a new lower
constraint on the radius from collisional mantle stripping (solid
line), corresponding to the lowest blue line in Figure
\ref{fig:mass_radius}, a head-on collision at 80~km~s$^{-1}$ between
equal-mass bodies.

\section{Discussion and Conclusions}

\begin{figure}
\figurenum{2}
\begin{center}
\includegraphics[scale=0.6]{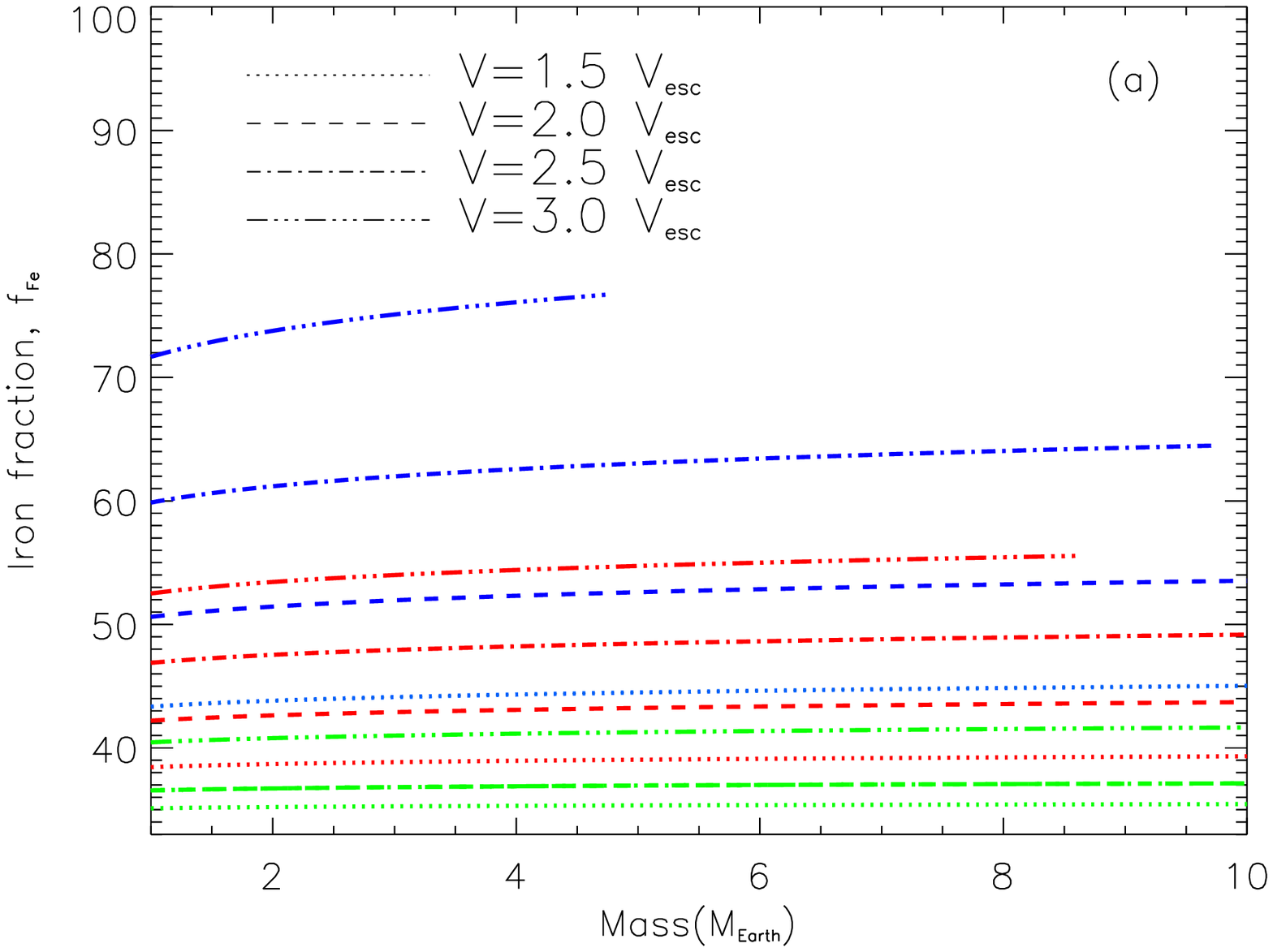}
\includegraphics[scale=0.6]{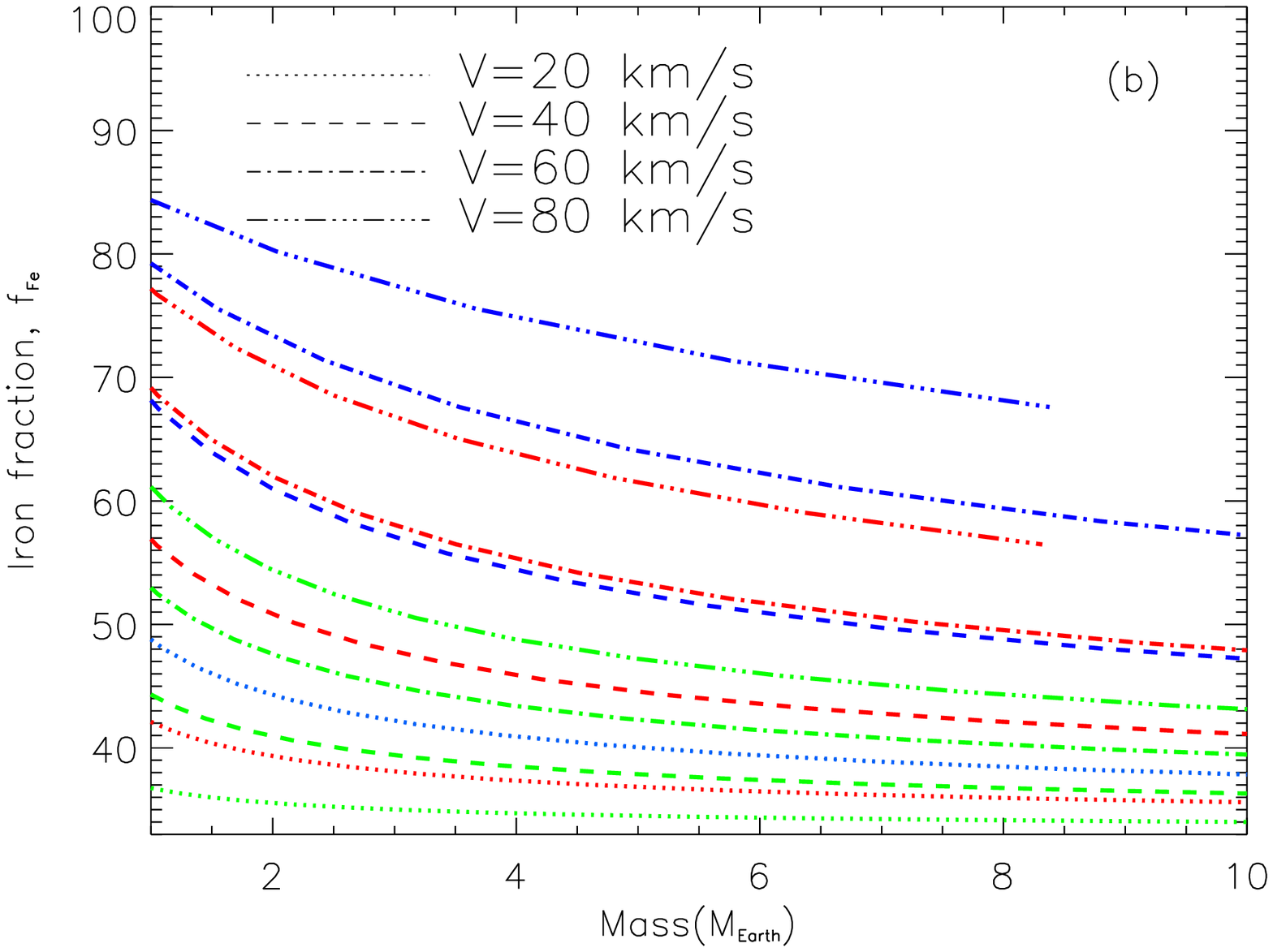}
\end{center}
\caption{Iron fraction of the largest remnant vs. mass of the largest
  remnant.  (a) Impact velocities given in terms of the mutual escape
  velocity. (b) Impact velocities in km~s$^{-1}$. In both panels, line
  color indicates the projectile-to-target mass ratio: green, 1/10;
  red, 1/4; blue, 1/1.  }
\label{fig:ironfrac}
\end{figure}

\begin{figure}
\figurenum{3}
\begin{center}
\includegraphics[scale=0.6]{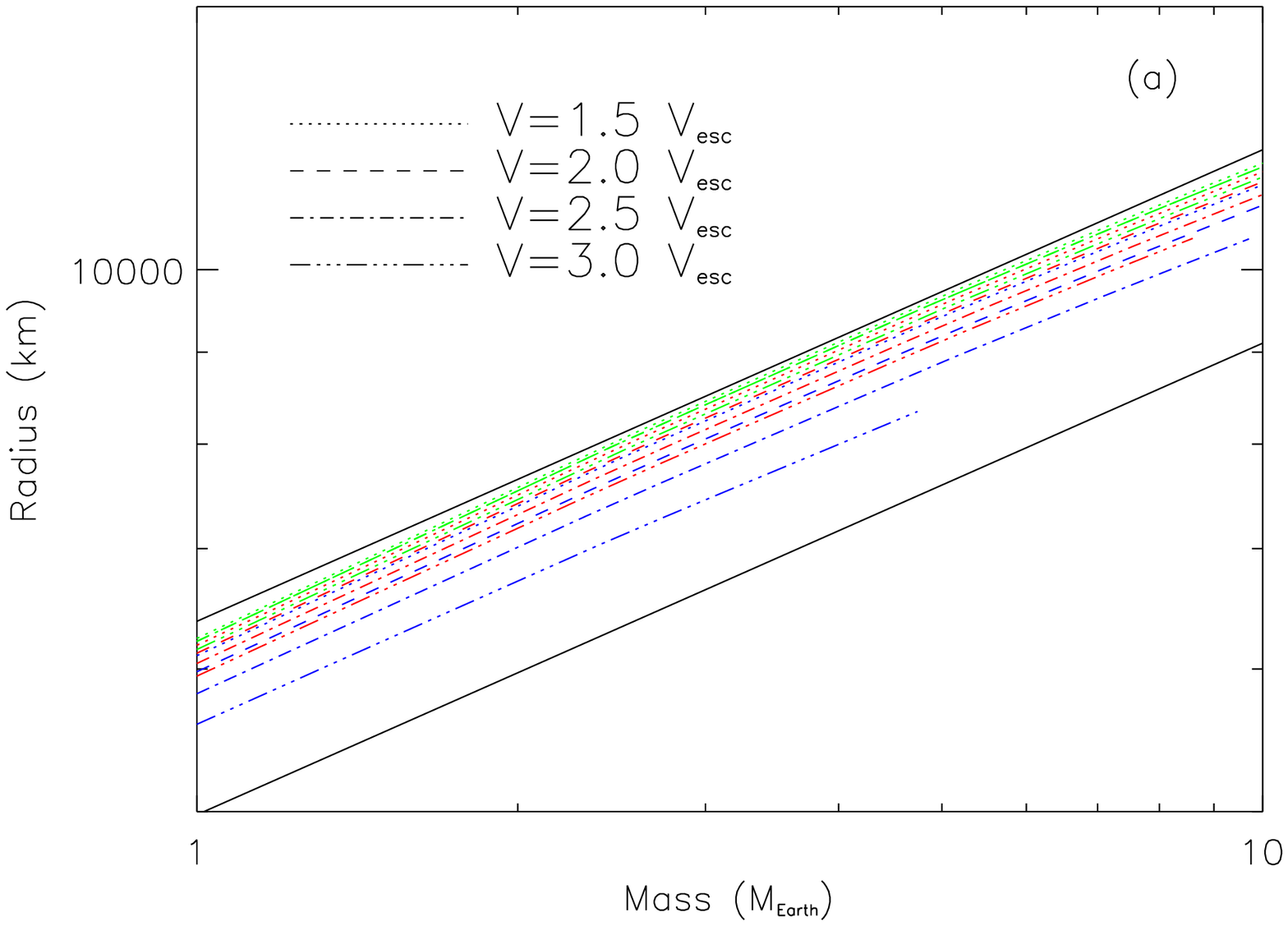}
\includegraphics[scale=0.6]{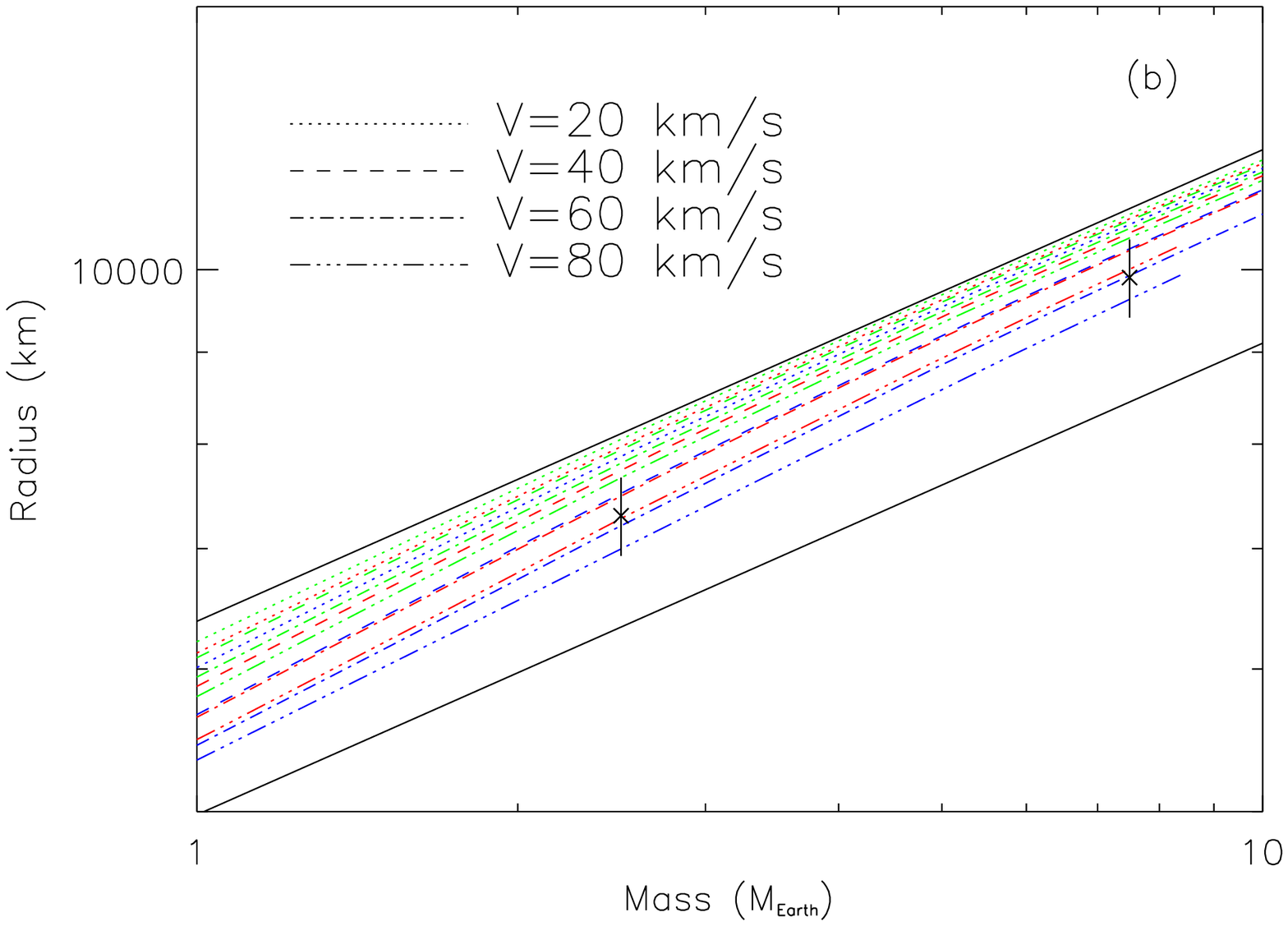}
\end{center}
\caption{Mass-radius diagram for super-Earths. (a) Impact velocities
  given in terms of the mutual escape velocity. (b) Impact velocities
  given in km~s$^{-1}$. In both panels, the two solid black lines
  represent terrestrial composition (iron mass fraction of 0.33, upper
  line) and pure iron (lower line) \citep{Valencia:2007b}.  The
  colored lines are for super-Earths stripped of mantle material in a
  giant impact.  Line color indicates the projectile-to-target mass
  ratio: green, 1/10; red, 1/4; blue, 1/1.  In (b), the $\times$
  symbols represent potentially observable transiting planets with 5\%
  uncertainty in the radii. Note that the difference in minimum radii
  between a pure iron planet and plausible densities achieved via
  collisional stripping is observable.}
\label{fig:mass_radius}
\end{figure}

\begin{figure}
\figurenum{4}
\begin{center}
\includegraphics[scale=0.6]{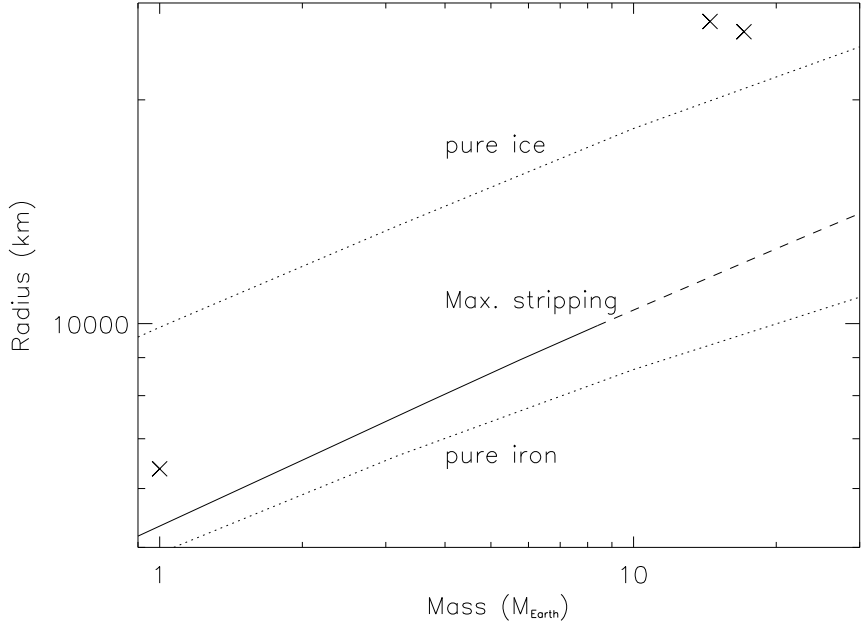}
\end{center}
\caption{Mass-radius diagram for super-Earths. The dotted lines are
  pure ice (upper) and pure iron (lower) \citep{Fortney:2007}.  The
  solid line is a constraint placed on the mass-radius diagram from
  collisional mantle stripping, corresponding to a 80~km~s$^{-1}$
  impact with a 1:1 projectile-to-target mass ratio. The extrapolation
  (dashed) corresponds to cases that require target masses
  $>15M_{\oplus}$. The $\times$ symbols indicate the masses and radii
  of Earth, Uranus, and Neptune.}
\label{fig:mass_radius2}
\end{figure}

Using the results of giant impact simulations presented in Marcus et
al. (2009), we have described the impact conditions necessary to strip
away mantle material from a nearly fully accreted planet. Combining
this with detailed interior structure models for super-Earths
\citep{Valencia:2006,Valencia:2007b, Fortney:2007}, we have
constructed a mass--radius diagram for super-Earths and shown that the
previous lower envelope in this relationship, corresponding to 100\%
iron super-Earths, is not consistent with collisional mantle
stripping.  Further, if the absence of super-Earths in the
10-100$M_{\oplus}$ range seen in the planet formation models of
\cite{Ida_LinI} is correct, even the existence of super-Mercuries,
$\sim$ 70\% iron by mass, may be limited to masses $\lesssim$
5$M_{\oplus}$ (as with the top blue curve in Figure
\ref{fig:ironfrac}(a)). This restriction arises because
$\sim10M_{\oplus}$ target bodies are required to make super-Mercuries
larger than about $5M_{\oplus}$.

The Kepler mission will discover a few hundred planets in the
mass--radius range of super-Earths shown in Figure 4. We predict that
the lower envelope of the distribution that Kepler is going to measure
will be significantly higher than our computed minimum based on
collisional stripping.  Our calculation derives a very conservative
limit with very low probability for these extreme scenario to be
realized. Hence, Kepler's limited sample of planets is unlikely to be
large enough to include such rare events. If super-Earths violating
the collisional stripping limit are actually confirmed around normal
stars, then we would need to revisit basic assumptions about the
planet formation process.

The simulations in this Letter were run on the Odyssey cluster
supported by the Harvard FAS Research Computing Group.

\clearpage

\clearpage

\clearpage

\clearpage

\bibliographystyle{apj}

\end{document}